\documentstyle[12pt]{article}
\newfont{\ffont}{msym10}                        
\newcommand{\beq}{\begin{equation}}             
\newcommand{\eeq}{\end{equation}}               
\newcommand{\bqry}{\begin{eqnarray}}            
\newcommand{\eqry}{\end{eqnarray}}              
\newcommand{\bqryn}{\begin{eqnarray*}}          
\newcommand{\eqryn}{\end{eqnarray*}}            
\newcommand{\NL}{\nonumber \\}                  
\newcommand{\preprint}[1]{\begin{table}[t]      
            \begin{flushright}                  
            \begin{large}{#1}\end{large}        
            \end{flushright}                    
            \end{table}}                        
\newcommand{\PD}[2]                             
    {\frac{\partial^{#2}}{\partial #1^{#2}}}    
\catcode`\@=11 \@addtoreset{equation}{section}  
\renewcommand{\theequation}                     
         {\arabic{section}.\arabic{equation}}   
\begin{document}
\preprint{TAUP-2180-94 \\  }
\title{Covariant Thermodynamics \\ and \\ ``Realistic'' Friedmann Model}
\author{\\ L. Burakovsky\thanks {Bitnet:BURAKOV@TAUNIVM.TAU.AC.IL.} \
and L.P. Horwitz\thanks {Bitnet:HORWITZ@TAUNIVM.TAU.AC.IL. Also at
Department of Physics, Bar-Ilan University, Ramat-Gan, Israel  } \\ \ }
\date{School of Physics and Astronomy \\ Raymond and Beverly Sackler
Faculty of Exact Sciences \\ Tel-Aviv University,
Tel-Aviv 69978, Israel}
\maketitle
\begin{abstract}
We discuss a cosmological Friedmann model modified by inclusion of
off-shell matter which has an equation of state $p,\rho \propto T^5,$
$p=1/4\rho .$ Such matter is shown to have energy density comparable with
that of non-interacting radiation at temperatures of the order of the
Hagedorn temperature, $\sim 10^{12}$ K, indicating the possibility of a
phase transition. It is argued that the $T^5$-phase, or an admixture,
lies below the high-temperature $T^4$-phase.
\end{abstract}

\bigskip
\section{Introduction}
The cosmological equations (without cosmological constant) for a
uniform universe with the space-time metric
\beq
-ds^2=-dt^2+R^2(t)\left[
\frac{dr^2}{1-kr^2}+r^2(d\theta ^2+\sin ^2\theta d\phi ^2)\right]
\eeq
are \cite{Wei}
\beq
\left( \frac{dR}{dt}\right) ^2=\frac{8\pi G}{3}\rho R^2-k,
\eeq
\beq
R^3\frac{dp}{dt}=\frac{d}{dt}\left[ R^3(\rho c^2+p)\right] ,
\eeq
where $R\equiv R(t)$ scales the comoving coordinates, $\rho (t)$ is the
energy density, $p(t)$ the pressure, $k=0,\pm 1$ the curvature constant
and $t$ is a universal cosmic time. The Friedmann models \cite{Fr} have
the properties: $p\simeq 0,\;k=0,\pm 1$ ($k=0$ is in fact the Einstein-de
Sitter model \cite{EdS}), and were improved and made more realistic by
Lema$\hat{{\rm i}}$tre \cite{L} who included non-interacting radiation.
If one denotes by subscripts $m$ and $r$ matter and radiation,
respectively, one finds from (1.3), for non-interacting matter and
radiation: $\rho _m\propto R^{-3},\;\rho _r\propto R^{-4},$ for
$p_m<<\rho _mc^2,\;p_r=\rho _rc^2/3.$ Eq. (1.2) becomes
\beq
\left( \frac{dR}{dt}\right) ^2=\alpha _rR^{-2}+\alpha _mR^{-1}-k,
\eeq
\beq
\alpha _r=8\pi G\rho _rR^4/3,\;\;\;\alpha _m=8\pi G\rho _mR^3/3.
\eeq
Eq. (1.4) is just the Lema$\hat{{\rm i}}$tre equation. Its solution for
$k=1$ was given by de Sitter \cite{dS}, Tolman \cite{T} and Alpher and
Herman \cite{AH}, for $k=0,\pm 1$ by Chernin \cite{Ch} and Cohen
\cite{Co}, and for $k=0$ by Jacobs \cite{J}. Cosmological models
containing both matter and radiation were also discussed by McIntosh
\cite{McI} and Harrison \cite{Har}.

For non-interacting matter and radiation, $\rho _r\propto T^4$ and hence
$T\propto R^{-1};$ and $\rho _m\propto n\propto T^3,$ where $n$ is the
mean matter number density. Therefore, the dimensionless quantity
\beq
\eta =n\left( \frac{\hbar c}{k_BT}\right) ^3=
n^{(0)}\left( \frac{\hbar c}{k_BT^{(0)}}\right) ^3
\eeq
is constant. Let us evaluate this quantity by assuming that we are
dealing with baryonic matter.

There are about $10^{57}$ nucleons in a typical star. There are about
$10^{11}$ galaxies in the universe, and each galaxy has about $10^{11}$
stars. Thus there are about $10^{79}$ baryons in the universe (in
comparison with $10^{89}$ photons, a number which is obtained by
thermodynamic arguments). The present size of the observable universe is
$10^{28}$ cm. The baryon number density is therefore given by $$n_b\sim
\frac{10^{79}}{\frac{4\pi }{3}(10^{28})^3}\sim 10^{-6}\;{\rm cm}^{-3},$$
which we take for $n^{(0)}.$ Taking $T^{(0)}\simeq 2.7$ K (the microwave
background), one obtains $\eta \sim 10^{-9}.$ Let $\beta =\rho _r^{
(0)}/\rho _m^{(0)}$ (zero superscript denotes the present epoch); then
because $\rho _rc^2\sim k_BT(k_BT/\hbar c)^3,$
\beq
\beta \sim \frac{k_BT^{(0)}}{\eta m_nc^2}\sim 10^{-3},
\eeq
where $m_n$ is the nucleon mass. Let the superscript $(1)$ denotes the
epoch where $\rho _r\simeq \rho _m;$ then
\beq
\frac{\alpha _r}{\alpha _m}=R^{(1)}=\beta R^{(0)},
\eeq
and $\rho _r^{(1)}=\rho _m^{(0)}\beta ^{-3}\sim 10^9\rho _m^{(0)}.$ Also,
$T^{(1)}=T^{(0)}/\beta \simeq 3000$ K, and when $T>T^{(1)},$ the energy
density of radiation exceeds that of matter.

The numerical values of the energy densities are \cite{W2}
\beq
\rho _m^{(0)}\sim 2\cdot 10^{-31}\;{\rm g\;cm}^{-3},\;\;
\;\rho _r^{(0)}\sim 4\cdot 10^{-34}\;{\rm g\;cm}^{-3},
\eeq
\beq
\rho _r^{(1)}\sim 10^{-22}\;{\rm g\;cm}^{-3}.
\eeq
It follows that the temperature dependence of the energy densities can
consistently have the form
\beq
\rho _m(T)\simeq 10^{-31}\left( \frac{T}{3}\right)^3{\rm g\;cm}^{-3},\;\;
\;\rho _r(T)\simeq 10^{-22}\left( \frac{T}{3\cdot 10^3}\right)^4{\rm
g\;cm}^{-3}.
\eeq

As remarked by Harrison \cite{Har}, when $T>T^{(2)}\sim m_ec^2/k_B\simeq
5\cdot 10^9$ K ($m_e$ is the electron mass), the Lema$\hat{{\rm i}}$tre
equation breaks down because of lepton and hadron pair production. At
temperatures $T\stackrel{>}{\sim }T ^{(2)}$ the interaction of the
hadrons is strong and has mainly a resonant character, the masses of the
resonances being comparable with the temperature. Under these conditions
hadronic matter is neither an ideal nor an  ultra-relativistic gas, and
can be well characterized by a resonance spectrum \cite{Hag}
\beq
\tau (m)\sim m^a\exp \;(m/T_0),
\eeq
where $a$ and $T_0$ are parameters. The statistical sum diverges
when $T>T_0,$ which indicates that the theory involves a limiting
temperature (the so-called ``hadronic boiling point'' of Hagedorn
\cite{Hag}), whose numerical value is found to be
\beq
T_0=m_\pi c^2/k_B\sim 10^{12}\;{\rm K}
\eeq
($m_\pi $ is the pion mass), and in the vicinity of $T_0$ particle
creation is so violent that it is impossible to exceed this temperature.
In the temperature range $T^{(2)}\stackrel{<}{\sim }T\stackrel{<}{\sim }
T_0$ the hadronic gas is thought of as being well described by the
``realistic'' equation of state suggested by Shuryak \cite{Sh}, $p,\rho
\propto T^6$ (which gives the value of the velocity of sound $c^2_s=dp/d
\rho =0.20),$ in agreement with some of the relevant experimental data
\cite{ZS,Gor,And}. There are indications that $c^2_s$ may, however, take
the values both more and less than 0.20, i.e., 0.25 \cite{Sh1,Sh2,Zh} or
0.17 \cite{And,SW,WW,OW}.

The existence of the limiting temperature $T_0$ suggests that there is
some phase transition. It also means that the Lema$\hat{{\rm i}}$tre
equation, although reasonably realistic, is a good approximation
throughout the lifetime of the universe $except$ for its earliest
moments.

It is clear that, in fact, the known resonances form an essentially
discrete set of states, and the well-known arguments applied above are
based on an approximate idealization that considers the envelope of these
resonances as a continuous mass spectrum. There is, however, a consistent
(proper time) formulation of a manifestly covariant statistical mechanics
\cite{HSP,HSS,di}, based on the ideas of Fock \cite{Fock} and
Stueckelberg \cite{St}, in which the four components of energy-momentum
are considered as independent degrees of freedom, permitting
fluctuations from the mass shell.

In the present paper we shall use this manifestly-covariant framework
(which we review briefly in the next section) as a model for the
description of phenomena which take place in hot hadronic matter. We
argue that the phase transition at the Hagedorn limiting temperature
can represent a phase transition from an off-shell sector of the theory,
in which the relativistic ensemble is described within the framework of
the manifestly covariant relativistic statistical mechanics mentioned
above, to an ultrarelativistic independent particle phase.
Recently we have studied thermodynamic properties of the off-shell phase
\cite{di,ind} and its possible consequences in hadronic physics
\cite{hadr}, astrophysics \cite{BHS}, and cosmology \cite{therm,5D}.

As shown in ref. \cite{hadr}, the behavior of hadronic matter below the
Hagedorn limiting temperature coincides with that of a system which
includes both particles and antiparticles, with the additional mass
potential \cite{HSP} $\mu _K\simeq 0;$ such a system, within the
covariant framework, is described by the equation of state corresponding
to Shuryak's ``realistic'' one. In the present paper we shall show that
off-shell matter, if included into the Friedmann-Lema$\hat{{\rm i}}$tre
model on an equal footing with baryonic matter and non-interacting
radiation, has energy density comparable to that of non-interacting
radiation at temperatures of the order of the Hagedorn temperature,
indicating the possibility of a phase transition from strongly
interacting (off-shell) phase to non-interacting one.

\section{Relativistic $N$-body system}
In the framework of a manifestly covariant relativistic statistical
mechanics, the dynamical evolution of a system of $N$ particles, for the
classical case, is governed by equations of motion that are of the form
of Hamilton equations for the motion of $N$ $events$ which generate the
space-time trajectories (particle world lines) as functions of a
continuous Poincar\'{e}-invariant parameter $\tau ,$ called the
historical time \cite{St,HP}. These events are characterized by their
positions $q^\mu =(t,{\bf q})$ and energy-momenta $p^\mu =(E,{\bf p})$ in
an $8N$-dimensional phase-space. For the quantum case, the system is
characterized by the wave function $\psi _\tau (q_1,q_2,\ldots ,q_N)\in
L^2(R^{4N}),$ with the measure $d^4q_1d^4q_2\cdots d^4q_N\equiv d^{4N}q$
$(q_i\equiv q_i^\mu ;\;\;\mu =0,1,2,3;\;\;i=1,2,\ldots ,N),$ describing
the distribution of events, which evolves with a generalized
Schr\"{o}dinger equation \cite{HP}. The collection of events (called
``concatenation'' \cite{AHL}) along each world line corresponds to a
{\it particle,} and hence, the evolution of the state of the $N$-event
system describes, {\it a posteriori,} the history in space and time of
an $N$-particle system.

For a system of $N$ interacting events (and hence, particles) one takes
\cite{HP} (we use the metric $g^{\mu \nu }=(-,+,+,+))$
\beq
K=\sum _i\frac{p_i^\mu p_{i\mu }}{2M}+V(q_1,q_2,\ldots ,q_N),
\eeq
where $M$ is a given fixed parameter (an intrinsic property of the
particles), with the dimension of mass, which we take to be the same for
all the particles of the system. The Hamilton equations are
$$\frac{dq_i^\mu }{d\tau }=\frac{\partial K}{\partial p_{i\mu }}=\frac{p_
i^\mu }{M},$$
\beq
\frac{dp_i^\mu }{d\tau }=-\frac{\partial K}{\partial q_{i\mu }}=-\frac{
\partial V}{\partial q_{i\mu }}.
\eeq
In the quantum theory, the generalized Schr\"{o}dinger equation
\beq
i\frac{\partial }{\partial \tau }\psi _\tau (q_1,q_2,\ldots ,q_N)=K
\psi _\tau (q_1,q_2,\ldots ,q_N)
\eeq
describes the evolution of the $N$-body wave function
$\psi _\tau (q_1,q_2,\ldots ,q_N).$

\subsection{Covariant thermodynamics and cosmology}
Thermodynamic functions for a many-body system can be derived from the
grand partition function, which, for an ensemble of off-shell events at
temperature $T$ is defined by the following expression,
modified, in comparison with the standard one, by the presence
of the term $\mu _KK$ (henceforth we use the system of units in which
$\hbar =c=k_B=1,$ unless the other specified):
\beq
Z=Tr\left[\exp \left\{-(\hat{E}^{(N)}-\mu \hat{N}-
\mu _K\hat{K}^{(N)})/T\right\}\right],
\eeq
where $\mu $ is the chemical potential and $\mu _K$ is the additional
mass potential of the ensemble \cite{HSP}.
Here $\hat{E}^{(N)}$ is the operator of the total energy of the $N$-body
system, $\hat{E}^{(N)}=\sum _{i=1}^Np^0_i,$ $\hat{K}^{(N)}$ is the
generalized $N$-body Hamiltonian defined by Eq. (2.1), and
$\hat{N}$ is the operator of the number of events (and therefore,
particles). As shown in ref. \cite{therm} (we shall review this point
briefly below), the quantity $-\mu _K\kappa ,$ where $\kappa $ is the
generalized Hamiltonian density, $\kappa =\langle \hat{K}^{(N)}\rangle /
V,$ represents a (55)-component of a generalized energy-momentum tensor,
which in the local rest frame takes on the form $T_{\alpha \beta }={\rm
diag}(-\rho ,p,p,p,-\sigma \mu _K\kappa ),\;\;\alpha ,\beta =0,1,2,3,5.$
Here the sign of $\sigma $ corresponds to the invariance group of the
extended $(x^\mu ,\tau )$ manifold which could be $O(4,1)$ $(\sigma =1)$
or $O(3,2)$ $(\sigma =-1.)$ In the present paper we shall restrict
ourselves to the case $\sigma =1,$ i.e., the $O(4,1)$ invariance of the
$(x^\mu ,\tau )$ manifold having the metric $g^{\alpha \beta }=
(-,+,+,+,+).$

Expressions for $p$ and $\rho ,$ using the grand canonical ensemble
obtained by Horwitz, Schieve and Piron \cite{HSP} in their study of
manifestly covariant statistical mechanics, were found in \cite{ind} in
terms of confluent hypergeometric functions as
\beq
p=\frac{T_{\triangle V}}{4\pi ^3}\frac{M^2}{\mu _K^3}T^3\sum _{s=1}^
\infty \frac{(\pm 1)^{s+1}}{s^3}e^{s\mu /T}\Psi (3,3;\frac{sM}{2\mu _KT})
,
\eeq
\beq
p+\rho =\frac{3T_{\triangle V}}{4\pi ^3}\frac{M^3}{\mu _K^4}T^2
\sum _{s=1}^\infty \frac{(\pm 1)^{s+1}}{s^2}e^{s\mu /T}\Psi (4,4;\frac{
sM}{2\mu _KT}),
\eeq
where $T_{\triangle V}$ is a characteristic interval of
$\tau $ for a trajectory to pass through a small representative
four-volume. For $T$ small, one finds from (2.5),(2.6) \cite{ind,therm}
that $p,\rho \propto T^6,$ $\rho \simeq 5p,$ and, in fact, that $\mu _K
\kappa \propto T^7$ is negligible in comparison with $\rho .$ On the
other hand, for $T$ large, it follows from these expressions that $p,\rho
,\mu _K\kappa \propto T^5,$ $p\simeq 1/4\rho \simeq -\mu _K\kappa .$ Now,
we remark that the energy-momentum tensor
\beq
T^{\mu \nu }=(p+\rho )u^\mu u^\nu -pg^{\mu \nu }
\eeq
can be extended to a five-dimensional form
\beq
^{(5)}T_{\alpha \beta }=\left(\;^{(5)}T_{\mu \nu },\;^{(5)}T_{55}\right);
\eeq
the requirement that the limiting case of the corresponding gravitational
theory (for zero curvature in the $\tau $ direction) coincide with the
Einstein equations results in the identification $^{(5)}T_{\mu \nu }=\;^{
(4)}T_{\mu \nu }$ and $^{(5)}T_{55}=-\mu _K\kappa. $ For high
temperature, it therefore follows that (as discussed in \cite{therm})
\beq
T^{\alpha \beta }=(p+\rho )u^\alpha u^\beta -pg^{\alpha \beta },\;\;\;
u^\lambda u_\lambda =-1,
\eeq
so that, in the local rest frame,
$T_{\alpha \beta }={\rm diag}\;(-\rho ,p,p,p,p).$

The first law of thermodynamics reads
\beq
dE=TdS-pdV+\mu _idN_i,
\eeq
where $S$ is the entropy, $V=V(t)$ (the universal cosmic time $t$
corresponds to the $\tau $ of (2.2) and (2.3)) is a comoving volume
element of an ideally uniform universe, and $N_i$ is the number of the
$i$th kind of particles having the chemical potential $\mu _i,$ which
implies that the energy $E$ is a thermodynamic function of $S,V,$ and $N_
i:$ $E=E(S,V,N_i).$ In the case of varying $N_i$ it is convenient to
consider the set of variables $(T,V,\mu _i)$ instead of $(S,V,N_i).$ The
change of variables is obtained with the help of a Legendre
transformation
\beq
\Omega (T,V,\mu _i)=E-TS-\mu _iN_i,
\eeq
which introduces the thermodynamic potential $\Omega .$ The use of (2.10)
in the differential of (2.11) leads to
\beq
d\Omega =-SdT-pdV-N_id\mu _i.
\eeq
The coefficients $S,p$ and $N_i$ are given by the partial derivatives
\beq
S=-\left(\frac{\partial \Omega }{\partial T}\right)_{V,\mu _i};\;\;\;
p=-\left(\frac{\partial \Omega }{\partial V}\right)_{T,\mu _i};\;\;\;
N_i=-\left(\frac{\partial \Omega }{\partial \mu _i}\right)_{T,V}.
\eeq
A fundamental result of statistical mechanics relates the thermodynamic
potential to the grand partition function:
\beq
\Omega (T,V,\mu _i)=-T\ln Z.
\eeq
By combining (2.13) with (2.14), the entropy and the pressure become
\bqry
S & = & \frac{\partial (T\ln Z)}{\partial T}, \\
p & = & \frac{\partial (T\ln Z)}{\partial V}.
\eqry
In the case of no interactions the latter relation reduces to
\beq
p=\frac{T}{V}\ln Z=-\frac{\Omega }{V},
\eeq
which is the equation of state of a free relativistic ensemble \cite{HSP}
\beq
\frac{pV}{T}=\ln Z.
\eeq
Substitution of $\Omega =-pV$ into Eq. (2.11) yields the thermodynamic
relation
\beq
E=-pV+TS+\mu _iN_i,
\eeq
which represents the formula for the entropy \cite{KT}
\beq
S=\frac{E+pV-\mu _iN_i}{T}=\frac{(\rho +p-\mu _iN_{0i})V}{T},
\eeq
where $\rho \equiv E/V,\;N_{0i}\equiv N_i/V$ are the corresponding
energy and particle number densities.

Since the lepton and baryon numbers are relatively small and therefore
negligible in the early universe, equilibrium conditions correspond to
\beq
\mu _i=0.
\eeq
Then the first law reads
\beq
d(\rho V)=TdS-pdV,
\eeq
and Eq. (2.20) takes on the form
\beq
S=\frac{(\rho +p)V}{T}.
\eeq
Since $V\sim R^3,$ it follows from (1.3) and (2.22) that
\beq
\frac{dS}{dt}=0,
\eeq
and the entropy is conserved. Note that Eq. (1.3) represents the local
energy conservation. Indeed, for the
energy-momentum tensor of an ideal cosmological fluid, $$T^{\mu \nu}=
(p+\rho )u^\mu u^\nu -pg^{\mu \nu },\;\;\;u^\rho u_\rho =-1,$$ the local
energy conservation, $\nabla _\mu T^{\mu \nu }=0,$ takes on the form
\beq
R\frac{d\rho }{dt}+3(\rho +p)\frac{dR}{dt}=0,
\eeq
which reduces to (1.3).

By rewriting (2.22) in the form
\beq
TdS=d\left[ (\rho +p)V\right] -Vdp
\eeq
and using the integrability condition \cite{KT}
\beq
\frac{\partial ^2S}{\partial T\partial V}=
\frac{\partial ^2S}{\partial V\partial T},
\eeq
the energy density and pressure are related as follows,
\beq
\rho +p=T\frac{dp}{dT}.
\eeq
Given an equation of state, the temperature dependence of $p$ and
$\rho $ can be derived with the help of (2.28).
Assuming that in the early universe the pressure is proportional to the
energy density and using Zeldovich's equation \cite{Z}
\beq
p=(\Gamma -1)\rho ,
\eeq
we find
\beq
p,\rho \propto T^{\Gamma /(\Gamma -1)}.
\eeq
In the standard framework one can obtain from (2.25),(2.29),
\beq
p,\rho \propto R^{-3\Gamma }.
\eeq
For the radiation-dominated universe $\Gamma =4/3,$ so that
\beq
p,\rho \propto T^4\propto R^{-4}.
\eeq
In the covariant framework Eq. (2.28), and therefore (2.30), hold, while
Eq. (2.25) should be modified. For the generalized energy-momentum tensor
(2.8) the local energy conservation, $\nabla _{\alpha }T^{\alpha \beta }
=0,$ yields
\beq
R\frac{d\rho }{dt}+4(\rho +p)\frac{dR}{dt}=0,
\eeq
resulting, via (2.29), in
\beq
p,\rho \propto R^{-4\Gamma }.
\eeq
Since in this case \cite{therm}
$\Gamma =5/4,$ we obtain from (2.30),(2.34)
\beq
p,\rho \propto T^5\propto R^{-5}.
\eeq

\section{``Realistic'' Friedmann model}
Let us modify the Friedmann-Lema$\hat{{\rm i}}$tre model by inclusion
of off-shell matter with the equation of state\footnote{The equation of
state of relativistic off-shell matter $p=1/4\rho $ was obtained also
by Hakim \cite{Hak} within a different framework.} $\rho _{m^{'}}\propto
T^5\propto R^{-5},\;p_{m^{'}}=\rho _{m^{'}}c^2/4.$ The modified equation
reads
\beq
\left( \frac{dR}{dt}\right) ^2=\alpha _{m^{'}}R^{-3}+\alpha _rR^{-2}+
\alpha _mR^{-1}-k,
\eeq
\beq
\alpha _{m^{'}}=8\pi G\rho _{m^{'}}R^5/3,
\eeq
$\alpha _r,\alpha _m$ being defined in (1.5).

Let $\beta ^{'}=\rho ^{(1)}_{m^{'}}/\rho ^{(1)}_r,$ where, as previously,
the superscript (1) denotes the epoch of $T^{(1)}\simeq 3000$ K. Because
$\rho _{m^{'}}c^2\sim k_BT(k_BT/\hbar c)^4,$
\beq
\beta ^{'}\sim \frac{k_BT^{(1)}}{\rho ^{(1)}_rc^2}=\frac{k_BT^{(0)}/
\beta }{\rho _m^{(0)}c^2/\beta ^3}=\beta ^2\frac{k_BT^{(0)}}{\rho _m^{
(0)}c^2}=\beta ^3\sim 10^{-9}.
\eeq
If superscript (3) denotes the epoch where $\rho _{m^{'}}\simeq \rho _r$
(implying that the pressure of radiation and off-shell matter are equal;
from (1.11) one sees that the contribution of baryonic matter is
negligible at high temperature), then
\beq
\frac{\alpha _{m^{'}}}{\alpha _r}=R^{(3)}=\beta ^{'}R^{(1)},
\eeq
and
\beq
\rho ^{(3)}_{m^{'}}=\rho ^{(1)}_r(\beta ^{'})^{-4}\sim 10^{36}\rho
^{(1)}_r\sim 10^{14}\;{\rm g\;cm}^{-3}.
\eeq
Also
\beq
T^{(3)}=\frac{T^{(1)}}{\beta ^{'}}\simeq 3\cdot 10^{12}\;{\rm K}.
\eeq
We see that this temperature is just of the order of the Hagedorn one.
At $T\simeq T^{(3)},$ off-shell matter and radiation have comparable
energy densities, indicating the possibility of a phase transition. In
general,
\beq
\rho _{m^{'}}(T)\simeq 10^{14}\left( \frac{T}{3\cdot 10^{12}}\right) ^5
{\rm g\;cm}^{-3}.
\eeq
Since the cosmological phase transition at $T_c\sim 10^{12}$ K is
normally associated with the transition from a strongly interacting
hadronic phase to a weakly interacting quark-gluon plasma phase
\cite{OW,GK}, we associate the $T^4$-phase above the transition
temperature with a phase of of weakly interacting quarks and gluons, and
$T^5$-phase below the transition temperature with a phase of strongly
interacting hadrons (as we show in \cite{realeq}, $T^4$ behavior indeed
results from the very high temperature asymptotic behavior of the
distribution function, for small $\mu _K.)$ In this phase, particles
undergo continual mutual interaction and are necessarily off-shell.
Therefore, the effect of strong interaction in such a system may be
represented by the off-shellness of its particles, justifying the use of
the off-shell framework for the description of this phase.

If the phase transition at temperature $\sim T^{(3)}$ is not
sufficiently sharp, there exists some region of the ``mixed'' phase, in
which both (or the three, including baryonic matter) phases coexist
(a similar situation may occur for both hadronic and the quark-gluon
plasma phases, according to the ``realistic'' scenario suggested by
Shuryak \cite{Sh}). If the off-shell phase extended well below $T^{(3)},$
it would have energy density comparable with that of baryonic matter at
$T\sim 10^8$ K, as follows from (1.11) and (3.7), so that the temperature
range of the mixed phase would be $10^8\;{\rm K}\stackrel{<}{\sim }T
\stackrel{<}{\sim }10^{12}\;{\rm K},$ where $\rho _m\stackrel{<}{\sim }
\rho _{m^{'}}\stackrel{<}{\sim }\rho _r.$ In fact, the mixed phase, as
we explain below, may exist in the temperature range
\beq
10^{10}\;{\rm K}\stackrel{<}{\sim }T\stackrel{<}{\sim }10^{12}\;{\rm K},
\eeq
where $10^{10}$ K is the electron-positron threshold for hadron
production in the $e^{+}e^{-}$ annihilation process, and at temperatures
well below $10^{10}$ K the off-shell phase is exponentially suppressed.
For $T$ from the range (3.8), we obtain, with the help of the relations
$$c^2=\frac{dp}{d\rho }=\frac{dp/dT}{d\rho /dT},$$
$$p=p_m+p_r+p_{m^{'}},\;\;\;\rho =\rho _m+\rho _r+\rho _{m^{'}},$$
the following expression for the velocity of sound:
\bqry
c^2 & = & \frac{n_mp_m+n_rp_r+n_{m^{'}}p_{m^{'}}}
{n_m\rho _m+n_r\rho _r+n_{m^{'}}\rho _{m^{'}}} \NL
 & = & \frac{n_mc_m^2+n_rc_r^2\rho _r/\rho _m+n_{m^{'}}c_{m^{'}}^2\rho _{
m^{'}}/\rho _m}{n_m+n_r\rho _r/\rho _m+n_{m^{'}}\rho _{m^{'}}/\rho _m},
\eqry
where $n_m,\;n_r,\;n_{m^{'}}$ are the powers of temperature
in the corresponding formulas for the pressure and the energy densities,
and $c_m^2=dp_m/d\rho _m=p_m/\rho _m,$ etc. are the sound velocities in
the corresponding phases. Using the relations $$c_m^2<<1,\;\;
\;c_r^2=\frac{1}{3},\;\;\;c_{m^{'}}^2=\frac{1}{4},$$ we finally obtain
\beq
c^2\simeq \frac{4/3\;\rho _r/\rho _m+5/4\;\rho _{m^{'
}}/\rho _m}{3+4\rho _r/\rho _m+5\rho _{m^{'}}/\rho _m}.
\eeq
If the energy densities of the corresponding phases were of the same
order, $\rho _m\sim \rho _r\sim \rho _{m^{'}},$ one would obtain
\beq
c^2\simeq \frac{31}{144}\approx 0.21.
\eeq
In fact, the use of Eqs. (1.11),(3.7) in the formula (3.10) provides for
the sound velocity a numerical value which is practically 1/3, up to
temperature $\sim 2\cdot 10^{12}$ K. At $T=T^{(3)}$ $(\simeq 3\cdot 10^{
12}$ K), Eq. (3.10) yields $c^2\approx 0.28.$ One sees that for the model
of the mixed phase, a dip in the sound velocity as a function of
temperature in the vicinity of the transition temperature is predicted,
in agreement with both the phenomenological models \cite{Sh,SZ} and the
recent data on QCD lattice simulations \cite{RS}. Let us also note that
the value of the sound velocity $c^2=0.28$ at $T\sim 10^{12}$ K was
reported in ref. \cite{Suh}.

The value of the sound velocity $c^2\approx 0.25,$ in agreement with the
relevant experimental data \cite{Sh1,Sh2,Zh}, is obtained only in the
assumption of a sufficiently sharp transition, when the phase below
the transition temperature is that of strongly interacting (off-shell)
matter alone, with temperature dependence $T^5.$

Thus, the ``realistic'' Friedmann model studied here describes
qualitatively the decrease of the sound velocity in relativistic gas at
values of $T$ in the vicinity of $10^{12}$ K, which reaches the numerical
value $c^2\approx 0.25,$ consistent with at least some of the
experimental results \cite{Sh1,Sh2,Zh}, and significantly less than
the ultra-relativistic Stefan-Boltzmann gas value 0.33. It may be thought
that the progressive population of states of higher mass creates an
admixture of heavy particles, thus decreasing the factor relating
pressure to energy density.

The above estimates have been made in a classical framework. Now we shall
show how similar results can be obtained within the framework of the
conventional (on-shell) quantum theory. As mentioned above, at $T^{(2)}
\sim m_e\simeq 5\cdot 10^9$ K, lepton and hadron pair production starts;
in the temperature range $T^{(2)}\stackrel{<}{\sim }T\stackrel{<}{\sim }
T^0,$ where $T_0$ is the temperature of the Hagedorn phase transition,
$T_0\sim T^{(3)},$ the system is characterized by a resonance spectrum
of the Hagedorn form (1.12), for which there is some experimental basis
\cite{Hag,Sh}. At $T\sim 10^{10}-10^{11}\;{\rm K}<<T_0,$ one can neglect
the exponential in (1.12) and use, as done by Shuryak \cite{Sh,S2}, a
resonance spectrum of the form
\beq
\tau (m)\sim m^a,
\eeq
where one chooses on phenomenological grounds $a=1$ \cite{Sh}.

The expressions for the pressure and energy density are then written as
\cite{Sh,S2} (we neglect the difference in properties of Bose and Fermi
particles at high temperature)
\bqry
p & = & \frac{g}{3}\int _{\triangle }^\infty \!\!\!dm\;\tau (m)\int \!
\frac{d^3k}{(2\pi )^3}\frac{k^2}{\sqrt{k^2+m^2}}e^{-\sqrt{k^2+m^2}/T}
\;=\;g\int _{\triangle }^\infty \!\!\!dm\;\tau (m)p_m, \\
\rho & = & g\int _{\triangle }^\infty \!\!\!dm\;\tau (m)\int \!
\frac{d^3k}{(2\pi )^3}\sqrt{k^2+m^2}e^{-\sqrt{k^2+m^2}/T}\;=\;g\int _
{\triangle }^\infty \!\!\!dm\;\tau (m)\rho _m,
\eqry
where $g$ is the number of hadronic degrees of freedom,
$p_m$ and $\rho _m$ are expressions for the pressure and energy
density of relativistic gas of particles with given mass $m,$ and we may
choose, sufficient for our present purposes, $\triangle \cong 2m_e\simeq
10^{10}$ K. Since at temperatures $T\stackrel{>}{\sim }\triangle $ one
should take into account electron-positron annihilation, leading, via
quark loops, to the formation of hadronic jets \cite{Per},
\beq
e^{+}e^{-}\rightarrow q\bar{q}\rightarrow {\rm hadrons,}
\eeq
which can be thought of as resonances on the excitation background of
virtual leptonic pair states, the $e^{+}e^{-}$ threshold should be taken
as the lowest possible mass for the resonance spectrum $\tau (m),$
justifying the introduction of the corresponding cut-off in the formulas
(3.13),(3.14).

The use of the standard expression \cite{KT1}
\beq
p_m=\frac{m^2T^2}{2\pi ^2}K_2\left( \frac{m}{T}\right)
\eeq
in Eq. (3.13), in which we take $\tau (m)=Cm,\;C={\rm const,}$ yields
\cite{GrRy}
\beq
p=\frac{gCT^2}{2\pi ^2}\int _{\triangle }^\infty dm\;m^3K_2\left(
\frac{m}{T}\right) =\frac{gCT^3}{2\pi ^2}\triangle ^3K_3\left( \frac{
\triangle }{T}\right) .
\eeq
At temperatures $T\sim 10^{11}$ K where $T>>\triangle $ (but
still $T<<T_0,$ so that the use of (3.12) instead of (1.12) is
justified), one uses the asymptotic formula \cite{AS}
\beq
K_\nu (z)\sim \frac{1}{2}\Gamma (\nu )\left( \frac{z}{2}\right) ^{-\nu },
\;\;\;z<<1
\eeq
and obtains
\beq
p=\frac{4gC}{\pi ^2}T^6.
\eeq
Similarly, one can obtain for such temperatures \cite{Sh},
\beq
\rho =\frac{20gC}{\pi ^2}T^6=5p,
\eeq
so that
\beq
c^2=\frac{dp}{d\rho }=0.20.
\eeq
Equations (3.19)-(3.21) represents the ``realistic'' equation of state
suggested by Shuryak for hot hadronic matter \cite{Sh}.

For $T<<\triangle ,$ one uses another asymptotic formula \cite{AS}:
\beq
K_\nu (z)\sim \sqrt{\frac{\pi }{2z}}e^{-z},\;\;\;z>>1
\eeq
and obtains
\beq
p=\frac{gCT^{7/2}\triangle ^{5/2}}{2^{5/2}\pi ^{3/2}}e^{-\triangle /T},
\eeq
and similar expression for $\rho .$ Thus, at temperatures well below
$\triangle \simeq 10^{10}$ K, the off-shell phase is suppressed by the
exponential.

The use of $\tau (m)\sim m^a$ in the formulas (3.13),(3.14) will
analogously lead to the equation of state at $T\sim 10^{11}$ K
\cite{Sh,S2}
\beq
p,\rho \sim T^{a+5},\;\;\;p=\rho /(a+4),
\eeq
which coincides with (3.19)-(3.21) for $a=1.$ The values for the sound
velocity in the hadronic phase 0.25 and 0.17, in agreement with some of
the experimental results \cite{And}-\cite{OW}, is achieved, in view of
(3.24), for $a=0$ and $a=2,$ respectively.

\newpage
\section{Solutions to cosmological equation}
The solutions of Eq. (3.1) for $t>t^{(3)},\;R>R^{(3)}$ are well known
\cite{Har},
\bqryn
k=0:  &   &  \\
R & = & \sqrt{\alpha _r}\tau +\frac{1}{4}\alpha _m\tau ^2, \\
t & = & \frac{1}{2}\sqrt{\alpha _r}\tau ^2+\frac{1}{12}\alpha _m\tau ^3;
\\
k=1:  &   &  \\
R & = & \sqrt{\alpha _r}\sin \tau +\alpha _m\sin ^2\frac{\tau }{2}, \\
t & = & 2\sqrt{\alpha _r}\sin ^2\frac{\tau }{2}+\frac{1}{2}\alpha _m(
\tau -\sin \tau); \\
k=-1:  &   &  \\
R & = & \sqrt{\alpha _r}\sinh \tau +\alpha _m\sinh ^2\frac{\tau }{2}, \\
t & = & 2\sqrt{\alpha _r}\sinh ^2\frac{\tau }{2}+\frac{1}{2}\alpha _m(
\sinh \tau -\tau).
\eqryn
For $0<t<t^{(3)},\;0<R<R^{(3)},$ since the curvature term is negligible
in the early universe, we have the equation of the general form
\beq
\left( \frac{dR}{dt}\right) ^2=\alpha _nR^{2-n},\;\;\;\alpha _n=8\pi G
\rho R^n/3,
\eeq
with $n=5.$ Eq. (4.1) has the solution
\beq
R=\alpha _n^{1/(n-2)}\left( \frac{n-2}{2}\tau \right) ^{2/(n-2)},
\eeq
\beq
t=\frac{2}{n}\alpha _n^{1/(n-2)}\left(\frac{n-2}{2}\tau \right)^{
n/(n-2)};
\eeq
so that
\beq
R=\left( \frac{n}{2}\right) ^{2/n}\alpha _n^{1/n}t^{2/n},
\eeq
and
\beq
t=\frac{2}{n}\left( \frac{8\pi G}{3}\rho \right)^{-1/2}.
\eeq
For the energy density at $T=T^{(3)},$ $\rho _{m^{'}}^{(3)}\sim
10^{14}\;{\rm g\;cm}^{-3},$ we obtain from (4.5), with $n=5,$
\beq
t^{(3)}\sim 10^{-5}{\rm c}.
\eeq
Equation (4.1) corresponds to that of Zeldovich and Novikov \cite{ZN}
for the early charged symmetric universe, whereas the
Lema$\hat{{\rm i}}$tre equation (1.4) is for the subsequent charge
asymmetric universe in which the asymmetry is either local or global.

\section{Concluding remarks}
We have studied the cosmological Friedmann model modified by introduction
of off-shell matter with the equation of state $\rho \propto T^5\propto
R^{-5},\;p=\rho /4.$ The energy density of such a matter is comparable
with that of the non-interacting radiation at temperature of the order of
the Hagedorn limiting temperature, $\sim 10^{12}$ K, indicating the
possibility of a phase transition.

A cosmological phase transition at $T_c\sim 10^{12}$ K is normally
associated with the transition from a strongly interacting hadronic phase
to a weakly interacting quark-gluon plasma phase \cite{OW,GK}. The
simplest classical model considered in the present paper implies the
possibility of a first order phase transition from the $T^5$-phase of
strongly interacting (off-shell) matter to the $T^4$-phase of
non-interacting radiation-like matter. Cosmological consequences of such
a phase transition are discussed in ref. \cite{5D}, where it is suggested
that the transition may be sufficiently smooth (second order) to preserve
the expansion rate. Although a first order phase transition might be
preferable for some cosmological implications, due to the fluctuations
which are generated at the transition and could produce planetary mass
black holes \cite{CS} which, in turn, could provide a possible
explanation for the dark matter of the universe and even be seeds in
galaxy formation \cite{FPS,Wi}, experimental
indications on the order of this phase transition are still absent.
Indeed, presently available lattice data on $SU(N)$ pure gauge theory
lattice simulations indicate that a phase transition to a weakly
interacting phase is of apparently first order for $SU(3)$ and second
order for $SU(2)$ theory \cite{Mul}. In ref. \cite{Br}, however, the
apparent first order nature of the transition in the case of $SU(3)$ pure
gauge theory has been called in question. Moreover, there are indications
from lattice QCD calculations that when fermions are included, the phase
transition may be of second or higher order \cite{Ben}. In this case, as
remarked by Ornik and Weiner \cite{OW}, the phase transition would be
hardly distinguishable from a situation in which no phase transition
would have taken place (radiation-dominated universe alone).

\section*{Acknowledgements}
We wish to thank J. Eisenberg for a useful discussion.


\newpage
\begin{thebibliography}{9}
\bibitem{Wei} S. Weinberg, {\it Gravitation and Cosmology,}
(Wiley, New York, 1972), p. 472
\bibitem{Fr} A. Friedmann, Z. Phys. {\bf 10} (1922) 377, {\bf 21} (1924)
326
\bibitem{EdS} A. Einstein and W. de Sitter, Proc. Nat. Acad. Sci.
{\bf 18} (1932) 213
\bibitem{L} G. Lema$\hat{{\rm i}}$tre, Ann. Soc. Sci. Brux. {\bf 47 A}
(1927) 49; Bull. Astr. Insts. Neth. {\bf 5} (1930) 200; Mon. Not. R.
Astr. Soc. {\bf 95} (1931) 483
\bibitem{dS} W. de Sitter, Bull. Astr. Insts. Neth. {\bf 5} (1930) 193
\bibitem{T} R.C. Tolman, {\it Relativity, Thermodynamics and Cosmology,}
(Dover, New York, 1987), pp. 407-411
\bibitem{AH} R.A. Alpher and R.C. Herman, Nature Lond. {\bf 142} (1948)
419
\bibitem{Ch} A.D. Chernin, Soviet Astr. {\bf 9} (1966) 871
\bibitem{Co} J.M. Cohen, Nature Lond. {\bf 216} (1967) 249
\bibitem{J} K.C. Jacobs, Nature Lond. {\bf 215} (1967) 1157
\bibitem{McI} C.B.G. McIntosh, Mon. Not. R. Astr. Soc. {\bf 138} (1968)
423
\bibitem{Har} E.R. Harrison, Mon. Not. R. Astr. Soc. {\bf 140} (1968) 281
\bibitem{W2} Ref. [1], Chapter 15, Section 6
\bibitem{Hag} R. Hagedorn, Nuovo Cimento {\bf 35} (1965) 216; Suppl.
Nuovo Cimento {\bf 3} (1965) 147, {\bf 6} (1968) 311
\bibitem{Sh} E.V. Shuryak, {\it The QCD Vacuum, Hadrons and the
Superdense Matter,} (World Scientific, Singapore, 1988)
\bibitem{ZS} O.V. Zhirov and E.V. Shuryak, Sov. J. Nucl. Phys. {\bf 21}
(1975) 443
\bibitem{Gor} M.I. Gorenstein {\it et al.,} Phys. Lett. B {\bf 60} (1976)
283
\bibitem{And} B. Andersson {\it et al.,} Nucl. Phys. B {\bf 112} (1976)
413
\bibitem{Sh1} E.V. Shuryak, Sov. J. Nucl. Phys. {\bf 24} (1976) 330
\bibitem{Sh2} E.V. Shuryak, Phys. Lett. B {\bf 78} (1978) 150
\bibitem{Zh} O.V. Zhirov, Sov. J. Nucl. Phys. {\bf 30} (1979) 571
\bibitem{SW} S. Sohlo and G. Wilk, Lett. Nuovo Cimento {\bf 13} (1975)
375
\bibitem{WW} K. Wehrberger and R.M. Weiner, Phys. Rev. D {\bf 31} (1985)
222
\bibitem{OW} U. Ornik and R.M. Weiner, Phys. Rev. D {\bf 36} (1987) 1263
\bibitem{HSP} L.P. Horwitz, W.C. Schieve and C. Piron,  Ann. Phys.
{\bf 137} (1981) 306
\bibitem{HSS} L.P. Horwitz, S. Shashoua and W.C. Schieve, Physica A
{\bf 161} (1989) 300
\bibitem{di} L. Burakovsky and L.P. Horwitz, Physica A {\bf 201} (1993)
666
\bibitem{Fock} V.A. Fock, Phys. Z. Sowjetunion {\bf 12} (1937) 404
\bibitem{St} E.C.G. Stueckelberg, Helv. Phys. Acta {\bf 14} (1941) 372,
588; {\bf 15} (1942) 23
\bibitem{ind} L. Burakovsky and L.P. Horwitz, Found. Phys. {\bf 25}
(1995) 785
\bibitem{hadr} L. Burakovsky and L.P. Horwitz, Found. Phys., {\it in
press}
\bibitem{BHS} L. Burakovsky, L.P. Horwitz and W.C. Schieve, Statistical
Mechanics of Relativistic Degenerate Fermi Gas I.Cold Adiabatic Equation
of State, Preprint TAUP-2124-93
\bibitem{therm} L. Burakovsky and L.P. Horwitz, Manifestly Covariant
Relativistic Thermodynamics and Avoidance of Gravitational Singularities,
Preprint TAUP-2173-94
\bibitem{5D} L. Burakovsky and L.P. Horwitz, 5D Generalized Inflationary
Cosmology, to be published in Gen. Rel. Grav.
\bibitem{HP} L.P. Horwitz and C. Piron, Helv. Phys. Acta {\bf 46} (1973)
316
\bibitem{AHL} R. Arshansky, L.P. Horwitz and Y. Lavie, Found Phys.
{\bf 13} (1983) 1167
\bibitem{KT} E.W. Kolb and M.S. Turner, {\it The Early Universe,}
(Addison-Wesley, Redwood, CA, 1990), p. 66
\bibitem{Z} Y.B. Zeldovich, Sov. Phys. JETP {\bf 14} (1962) 1143
\bibitem{Hak} R. Hakim, J. Math. Phys. {\bf 15} (1974) 1310
\bibitem{GK} T. DeGrand and K. Kajantie, Phys. Lett. B {\bf 147} (1984)
273
\bibitem{realeq} L. Burakovsky, L.P. Horwitz and W.C. Schieve, Towards a
Realistic Equation of State of Strongly Interacting Matter, Preprint
TAUP-2239-95
\bibitem{SZ} E.V. Shuryak and O.V. Zhirov, Phys. Lett. B {\bf 89} (1980)
253
\bibitem{RS} T. {\c C}elik, J. Engels and H. Satz, Phys. Lett. B
{\bf 129} (1983) 323; \\ K. Redlich and H. Satz, Phys. Rev. D {\bf 33}
(1986) 3747
\bibitem{Suh} E. Suhonen {\it et al.,} Phys. Rev. Lett. {\bf 31} (1973)
1567
\bibitem{S2} E.V. Shuryak, Sov. J. Nucl. Phys. {\bf 16} (1973) 220
\bibitem{Per} D.H. Perkins, {\it Introduction to High Energy Physics,}
(Addison-Wesley, Reading, MA, 1982), p. 305
\bibitem{KT1} Ref. [37], p. 61
\bibitem{GrRy} I.S. Gradshteyn and I.M. Ryzhik, {\it Tables of Integrals,
Series, and Products,} (Academic Press, New York, 1980), subsection 6.561
\bibitem{AS} M. Abramowitz and I.A. Stegun, {\it Handbook of Mathematical
Functions,} (Dover, New York, 1972), p. 375
\bibitem{ZN} Y.B. Zeldovich and I.D. Novikov, Sov. Phys. JETP Lett.
{\bf 4} (1966) 80
\bibitem{CS} M. Crawford and D.N. Schramm, Nature {\bf 298} (1982) 538
\bibitem{FPS} K. Freese, R. Price and D.N. Schramm, Astrophys. J.
{\bf 275} (1983) 405
\bibitem{Wi} E. Witten, Phys. Rev. D {\bf 30} (1984) 272
\bibitem{Mul} B. M\"{u}ller, {\it The Physics of the Quark-Gluon Plasma,}
(Springer-Verlag, Berlin, 1985), p. 66
\bibitem{Br} F.R. Brown {\it et al.,} Phys. Rev. Lett. {\bf 61} (1988)
2058
\bibitem{Ben} B. Svetitsky, Nucl. Phys. A {\bf 461} (1987) 71c
\end{thebibliography}
\end{document}